# Measuring the muon content of inclined air showers using AERA and the water-Cherenkov detectors of the Pierre Auger Observatory

A. Abdul Halim,[13] P. Abreu,[70] M. Aglietta,[53,51] I. Allekotte,[1] K. Almeida Cheminant,[78,77] A. Almela,[7,12] R. Aloisio,[44,45] J. Alvarez-Muñiz,[76] A. Ambrosone,[44] J. Ammerman Yebra,[76] G. A. Anastasi,[57,46] L. Anchordoqui,[83] B. Andrada,[7] L. Andrade Dourado,[44,45] S. Andringa,[70] L. Apollonio,[58,48] C. Aramo,[49] E. Arnone,[62,51] J. C. Arteaga Velázquez,[66] P. Assis,[70] G. Avila,[11] E. Avocone,[56,45] A. Bakalova,[31] F. Barbato,[44,45] A. Bartz Mocellin,[82] J. A. Bellido,[13] C. Berat,[35] M. E. Bertaina,[62,51] M. Bianciotto,[62,51] P. L. Biermann,[a] V. Binet,[5] K. Bismark,[38,7] T. Bister,[77,78] J. Biteau,[36,i] J. Blazek,[31] J. Blümer,[40] M. Boháčová,[31] D. Boncioli,[56,45] C. Bonifazi,[8] L. Bonneau Arbeletche,[22] N. Borodai,[68] J. Brack,[f] P. G. Brichetto Orchera,[7,40] F. L. Briechle,[41] A. Bueno,[75] S. Buitink,[15] M. Buscemi,[46,57] M. Büsken,[38,7] A. Bwembya,[77,78] K. S. Caballero-Mora,[65] S. Cabana-Freire,[76] L. Caccianiga,[58,48] F. Campuzano,[6] J. Caraça-Valente,[82] R. Caruso,[57,46] A. Castellina,[53,51] F. Catalani,[19] G. Cataldi,[47] L. Cazon,[76] M. Cerda,[10] B. Čermáková,[40] A. Cermenati,[44,45] J. A. Chinellato,[22] J. Chudoba,[31] L. Chytka,[32] R. W. Clay,[13] A. C. Cobos Cerutti,[6] R. Colalillo,[59,49] R. Conceição,[70] G. Consolati,[48,54] M. Conte,[55,47] F. Convenga,[44,45] D. Correia dos Santos,[27] P. J. Costa,[70] C. E. Covault,[81] M. Cristinziani,[43] C. S. Cruz Sanchez,[3] S. Dasso,[4,2] K. Daumiller,[40] B. R. Dawson,[13] R. M. de Almeida,[27] E.-T. de Boone,[43] B. de Errico,[27] J. de Jesús,[7] S. J. de Jong,[77,78] J. R. T. de Mello Neto,[27] I. De Mitri,[44,45] J. de Oliveira,[18] D. de Oliveira Franco,[42] F. de Palma,[55,47] V. de Souza,[20] E. De Vito,[55,47] A. Del Popolo,[57,46] O. Deligny,[33] N. Denner,[31] L. Deval,[53,51] A. di Matteo,[51] C. Dobrigkeit,[22] J. C. D'Olivo,[67] L. M. Domingues Mendes,[16,70] Q. Dorosti,[43] J. C. dos Anjos,[16] R. C. dos Anjos,[26] J. Ebr,[31] F. Ellwanger,[40] R. Engel,[38,40] I. Epicoco,[55,47] M. Erdmann,[41] A. Etchegoyen,[7,12] C. Evoli,[44,45] H. Falcke,[77,79,78] G. Farrar,[85] A. C. Fauth,[22] T. Fehler,[43] F. Feldbusch,[39] A. Fernandes,[70] M. Fernandez,[14] B. Fick,[84] J. M. Figueira,[7] P. Filip,[38,7] A. Filipčič,[74,73] T. Fitoussi,[40] B. Flaggs,[87] T. Fodran,[77] A. Franco,[47] M. Freitas,[70] T. Fujii,[86,h] A. Fuster,[7,12] C. Galea,[77] B. García,[6] C. Gaudu,[37] P. L. Ghia,[33] U. Giaccari,[47] F. Gobbi,[10] F. Gollan,[7] G. Golup,[1] M. Gómez Berisso,[1] P. F. Gómez Vitale,[11] J. P. Gongora,[11] J. M. González,[1] N. González,[7] D. Góra,[68] A. Gorgi,[53,51] M. Gottowik,[40] F. Guarino,[59,49] G. P. Guedes,[23] L. Gülzow,[40] S. Hahn,[38] P. Hamal,[31] M. R. Hampel,[7] P. Hansen,[3] V. M. Harvey,[13] A. Haungs,[40] T. Hebbeker,[41] C. Hojvat,[d] J. R. Hörandel,[77,78] P. Horvath,[32] M. Hrabovský,[32] T. Huege,[40,15] A. Insolia,[57,46] P. G. Isar,[72] M. Ismaiel,[77,78] P. Janecek,[31] V. Jilek,[31] K.-H. Kampert,[37] B. Keilhauer,[40] A. Khakurdikar,[77] V. V. Kizakke Covilakam,[7,40] H. O. Klages,[40] M. Kleifges,[39] J. Köhler,[40] F. Krieger,[41] M. Kubatova,[31] N. Kunka,[39] B. L. Lago,[17] N. Langner,[41] N. Leal,[7] M. A. Leigui de Oliveira,[25] Y. Lema-Capeans,[76] A. Letessier-Selvon,[34] I. Lhenry-Yvon,[33] L. Lopes,[70] J. P. Lundquist,[73] M. Mallamaci,[60,46] D. Mandat,[31] P. Mantsch,[d] F. M. Mariani,[58,48] A. G. Mariazzi,[3] I. C. Mariş,[14] G. Marsella,[60,46] D. Martello,[55,47] S. Martinelli,[40,7] M. A. Martins,[76] H.-J. Mathes,[40] J. Matthews,[g] G. Matthiae,[61,50] E. Mayotte,[82] S. Mayotte,[82] P. O. Mazur,[d] G. Medina-Tanco,[67] J. Meinert,[37] D. Melo,[7] A. Menshikov,[39] C. Merx,[40] S. Michal,[31] M. I. Micheletti,[5] L. Miramonti,[58,48] M. Mogarkar,[68] S. Mollerach,[1] F. Montanet,[35] L. Morejon,[37] K. Mulrey,[77,78] R. Mussa,[51] W. M. Namasaka,[37] S. Negi,[31] L. Nellen,[67] K. Nguyen,[84] G. Nicora,[9] M. Niechciol,[43] D. Nitz,[84] D. Nosek,[30] A. Novikov,[87] V. Novotny,[30] L. Nožka,[32] A. Nucita,[55,47] L. A. Núñez,[29] J. Ochoa,[7,40] C. Oliveira,[20] L. Östman,[31] M. Palatka,[31] J. Pallotta,[9] S. Panja,[31] G. Parente,[76] T. Paulsen,[37] J. Pawlowsky,[37] M. Pech,[31] J. Pękala,[68] R. Pelayo,[64] V. Pelgrims,[14] L. A. S. Pereira,[24] E. E. Pereira Martins,[38,7] C. Pérez Bertolli,[7,40] L. Perrone,[55,47] S. Petrera,[44,45] C. Petrucci,[56] T. Pierog,[40] M. Pimenta,[70] M. Platino,[7] B. Pont,[77] M. Pourmohammad Shahvar,[60,46] P. Privitera,[86] C. Priyadarshi,[68] M. Prouza,[31] K. Pytel,[69] S. Querchfeld,[37] J. Rautenberg,[37] D. Ravignani,[7] J. V. Reginatto Akim,[22] A. Reuzki,[41] J. Ridky,[31] F. Riehn,[76,j] M. Risse,[43] V. Rizi,[56,45] E. Rodriguez,[7,40] G. Rodriguez Fernandez,[50] J. Rodriguez Rojo,[11] S. Rossoni,[42] M. Roth,[40] E. Roulet,[1] A. C. Rovero,[4] A. Saftoiu,[71] M. Saharan,[77] F. Salamida,[56,45] H. Salazar,[63] G. Salina,[50] P. Sampathkumar,[40] N. San Martin,[82] J. D. Sanabria Gomez,[29] F. Sánchez,[7] E. M. Santos,[21] E. Santos,[31] F. Sarazin,[82] R. Sarmento,[70] R. Sato,[11] P. Savina,[44,45] V. Scherini,[55,47] H. Schieler,[40] M. Schimassek,[33] M. Schimp,[37] D. Schmidt,[40] O. Scholten,[15,b] H. Schoorlemmer,[77,78] P. Schovánek,[31] F. G. Schröder,[87,40] J. Schulte,[41] T. Schulz,[31] S. J. Sciutto,[3] M. Scornavacche,[7] A. Sedoski,[7] A. Segreto,[52,46] S. Sehgal,[37] S. U. Shivashankara,[73] G. Sigl,[42] K. Simkova,[15,14] F. Simon,[39] R. Šmída,[86] P. Sommers,[e] R. Squartini,[10] M. Stadelmaier,[40,48,58] S. Stanič,[73] J. Stasielak,[68] P. Stassi,[35] S. Strähnz,[38] M. Straub,[41] T. Suomijärvi,[36] A. D. Supanitsky,[7] Z. Svozilikova,[31] K. Syrokvas,[30] Z. Szadkowski,[69] F. Tairli,[13] M. Tambone,[59,49] A. Tapia,[28] C. Taricco,[62,51] C. Timmermans,[78,77] O. Tkachenko,[31] P. Tobiska,[31] C. J. Todero Peixoto,[19] B. Tomé,[70] A. Travaini,[10] P. Travnicek,[31] M. Tueros,[3] M. Unger,[40] R. Uzeiroska,[37] L. Vaclavek,[32] M. Vacula,[32] I. Vaiman,[44,45] J. F. Valdés Galicia,[67] L. Valore,[59,49] P. van Dillen,[77,78] E. Varela,[63] V. Vašíčková,[37] A. Vásquez-Ramírez,[29] D. Veberič,[40] I. D. Vergara Quispe,[3] S. Verpoest,[87] V. Verzi,[50] J. Vicha,[31] J. Vink,[80] S. Vorobiov,[73] J. B. Vuta,[31] C. Watanabe,[27] A. A. Watson,[c] A. Weindl,[40] M. Weitz,[37] L. Wiencke,[82] H. Wilczyński,[68] B. Wundheiler,[7] B. Yue,[37] A. Yushkov,[31] E. Zas,[76] D. Zavrtanik,[73,74] and M. Zavrtanik[74,73]

(The Pierre Auger Collaboration)*

[1]Centro Atómico Bariloche and Instituto Balseiro (CNEA-UNCuyo-CONICET), San Carlos de Bariloche, Argentina
[2]Departamento de Física and Departamento de Ciencias de la Atmósfera y los Océanos, FCEyN, Universidad de Buenos Aires and CONICET, Buenos Aires, Argentina
[3]IFLP, Universidad Nacional de La Plata and CONICET, La Plata, Argentina
[4]Instituto de Astronomía y Física del Espacio (IAFE, CONICET-UBA), Buenos Aires, Argentina
[5]Instituto de Física de Rosario (IFIR)—CONICET/U.N.R. and Facultad de Ciencias Bioquímicas y Farmacéuticas U.N.R., Rosario, Argentina
[6]Instituto de Tecnologías en Detección y Astropartículas (CNEA, CONICET, UNSAM), and Universidad Tecnológica Nacional—Facultad Regional Mendoza (CONICET/CNEA), Mendoza, Argentina
[7]Instituto de Tecnologías en Detección y Astropartículas (CNEA, CONICET, UNSAM), Buenos Aires, Argentina
[8]International Center of Advanced Studies and Instituto de Ciencias Físicas, ECyT-UNSAM and CONICET, Campus Miguelete—San Martín, Buenos Aires, Argentina
[9]Laboratorio Atmósfera—Departamento de Investigaciones en Láseres y sus Aplicaciones—UNIDEF (CITEDEF-CONICET), Argentina
[10]Observatorio Pierre Auger, Malargüe, Argentina
[11]Observatorio Pierre Auger and Comisión Nacional de Energía Atómica, Malargüe, Argentina
[12]Universidad Tecnológica Nacional—Facultad Regional Buenos Aires, Buenos Aires, Argentina
[13]University of Adelaide, Adelaide, S.A., Australia
[14]Université Libre de Bruxelles (ULB), Brussels, Belgium
[15]Vrije Universiteit Brussels, Brussels, Belgium
[16]Centro Brasileiro de Pesquisas Fisicas, Rio de Janeiro, RJ, Brazil
[17]Centro Federal de Educação Tecnológica Celso Suckow da Fonseca, Petropolis, Brazil
[18]Instituto Federal de Educação, Ciência e Tecnologia do Rio de Janeiro (IFRJ), Brazil
[19]Universidade de São Paulo, Escola de Engenharia de Lorena, Lorena, SP, Brazil
[20]Universidade de São Paulo, Instituto de Física de São Carlos, São Carlos, SP, Brazil
[21]Universidade de São Paulo, Instituto de Física, São Paulo, SP, Brazil
[22]Universidade Estadual de Campinas (UNICAMP), IFGW, Campinas, SP, Brazil
[23]Universidade Estadual de Feira de Santana, Feira de Santana, Brazil
[24]Universidade Federal de Campina Grande, Centro de Ciencias e Tecnologia, Campina Grande, Brazil
[25]Universidade Federal do ABC, Santo André, SP, Brazil
[26]Universidade Federal do Paraná, Setor Palotina, Palotina, Brazil
[27]Universidade Federal do Rio de Janeiro, Instituto de Física, Rio de Janeiro, RJ, Brazil
[28]Universidad de Medellín, Medellín, Colombia
[29]Universidad Industrial de Santander, Bucaramanga, Colombia
[30]Charles University, Faculty of Mathematics and Physics, Institute of Particle and Nuclear Physics, Prague, Czech Republic
[31]Institute of Physics of the Czech Academy of Sciences, Prague, Czech Republic
[32]Palacky University, Olomouc, Czech Republic
[33]CNRS/IN2P3, IJCLab, Université Paris-Saclay, Orsay, France
[34]Laboratoire de Physique Nucléaire et de Hautes Energies (LPNHE), Sorbonne Université, Université de Paris, CNRS-IN2P3, Paris, France
[35]University Grenoble Alpes, CNRS, Grenoble Institute of Engineering Université Grenoble Alpes, LPSC-IN2P3, 38000 Grenoble, France
[36]Université Paris-Saclay, CNRS/IN2P3, IJCLab, Orsay, France
[37]Bergische Universität Wuppertal, Department of Physics, Wuppertal, Germany
[38]Karlsruhe Institute of Technology (KIT), Institute for Experimental Particle Physics, Karlsruhe, Germany
[39]Karlsruhe Institute of Technology (KIT), Institut für Prozessdatenverarbeitung und Elektronik, Karlsruhe, Germany
[40]Karlsruhe Institute of Technology (KIT), Institute for Astroparticle Physics, Karlsruhe, Germany
[41]RWTH Aachen University, III. Physikalisches Institut A, Aachen, Germany
[42]Universität Hamburg, II. Institut für Theoretische Physik, Hamburg, Germany
[43]Universität Siegen, Department Physik—Experimentelle Teilchenphysik, Siegen, Germany
[44]Gran Sasso Science Institute, L'Aquila, Italy
[45]INFN Laboratori Nazionali del Gran Sasso, Assergi (L'Aquila), Italy

[46]INFN, Sezione di Catania, Catania, Italy
[47]INFN, Sezione di Lecce, Lecce, Italy
[48]INFN, Sezione di Milano, Milano, Italy
[49]INFN, Sezione di Napoli, Napoli, Italy
[50]INFN, Sezione di Roma "Tor Vergata," Roma, Italy
[51]INFN, Sezione di Torino, Torino, Italy
[52]Istituto di Astrofisica Spaziale e Fisica Cosmica di Palermo (INAF), Palermo, Italy
[53]Osservatorio Astrofisico di Torino (INAF), Torino, Italy
[54]Politecnico di Milano, Dipartimento di Scienze e Tecnologie Aerospaziali, Milano, Italy
[55]Università del Salento, Dipartimento di Matematica e Fisica "E. De Giorgi," Lecce, Italy
[56]Università dell'Aquila, Dipartimento di Scienze Fisiche e Chimiche, L'Aquila, Italy
[57]Università di Catania, Dipartimento di Fisica e Astronomia "Ettore Majorana," Catania, Italy
[58]Università di Milano, Dipartimento di Fisica, Milano, Italy
[59]Università di Napoli "Federico II," Dipartimento di Fisica "Ettore Pancini," Napoli, Italy
[60]Università di Palermo, Dipartimento di Fisica e Chimica "E. Segrè," Palermo, Italy
[61]Università di Roma "Tor Vergata", Dipartimento di Fisica, Roma, Italy
[62]Università Torino, Dipartimento di Fisica, Torino, Italy
[63]Benemérita Universidad Autónoma de Puebla, Puebla, México
[64]Unidad Profesional Interdisciplinaria en Ingeniería y Tecnologías Avanzadas del Instituto Politécnico Nacional (UPIITA-IPN), México, D.F., México
[65]Universidad Autónoma de Chiapas, Tuxtla Gutiérrez, Chiapas, México
[66]Universidad Michoacana de San Nicolás de Hidalgo, Morelia, Michoacán, México
[67]Universidad Nacional Autónoma de México, México, D.F., México
[68]Institute of Nuclear Physics PAN, Krakow, Poland
[69]University of Łódź, Faculty of High-Energy Astrophysics,Łódź, Poland
[70]Laboratório de Instrumentação e Física Experimental de Partículas—LIP and Instituto Superior Técnico—IST, Universidade de Lisboa—UL, Lisboa, Portugal
[71]"Horia Hulubei" National Institute for Physics and Nuclear Engineering, Bucharest-Magurele, Romania
[72]Institute of Space Science, Bucharest-Magurele, Romania
[73]Center for Astrophysics and Cosmology (CAC), University of Nova Gorica, Nova Gorica, Slovenia
[74]Experimental Particle Physics Department, J. Stefan Institute, Ljubljana, Slovenia
[75]Universidad de Granada and C.A.F.P.E., Granada, Spain
[76]Instituto Galego de Física de Altas Enerxías (IGFAE), Universidade de Santiago de Compostela, Santiago de Compostela, Spain
[77]IMAPP, Radboud University Nijmegen, Nijmegen, The Netherlands
[78]Nationaal Instituut voor Kernfysica en Hoge Energie Fysica (NIKHEF), Science Park, Amsterdam, The Netherlands
[79]Stichting Astronomisch Onderzoek in Nederland (ASTRON), Dwingeloo, The Netherlands
[80]Universiteit van Amsterdam, Faculty of Science, Amsterdam, The Netherlands
[81]Case Western Reserve University, Cleveland, Ohio, USA
[82]Colorado School of Mines, Golden, Colorado, USA
[83]Department of Physics and Astronomy, Lehman College, City University of New York, Bronx, New York, USA
[84]Michigan Technological University, Houghton, Michigan, USA
[85]New York University, New York, New York, USA
[86]University of Chicago, Enrico Fermi Institute, Chicago, Illinois, USA
[87]University of Delaware, Department of Physics and Astronomy, Bartol Research Institute, Newark, Delaware, USA

[*]Contact author: spokespersons@auger.org; http://www.auger.org.
[a]Max-Planck-Institut für Radioastronomie, Bonn, Germany.
[b]Also at Kapteyn Institute, University of Groningen, Groningen, The Netherlands.
[c]School of Physics and Astronomy, University of Leeds, Leeds, United Kingdom.
[d]Fermi National Accelerator Laboratory, Fermilab, Batavia, Illinois, USA.
[e]Pennsylvania State University, University Park, Pennsylvania, USA.
[f]Colorado State University, Fort Collins, Colorado, USA.
[g]Louisiana State University, Baton Rouge, Louisiana, USA.
[h]Present address: Graduate School of Science, Osaka Metropolitan University, Osaka, Japan.
[i]Institut universitaire de France (IUF), France.
[j]Present address: Technische Universität Dortmund and Ruhr-Universität Bochum, Dortmund and Bochum, Germany.

We present a novel approach for assessing the muon content of air showers with large zenith angles on a combined analysis of their radio emission and particle footprint. We use the radiation energy reconstructed by the Auger engineering radio array (AERA) as an energy estimator and determine the muon number independently with the water-Cherenkov detector array of the Pierre Auger Observatory, deployed on a 1500 m grid. We focus our analysis on air showers with primary energy above 4 EeV to ensure full detection efficiency. Over approximately ten years of accumulated data, we identify a set of 40 high-quality events that are used in the analysis. The estimated muon contents in data are compatible with those for iron primaries as predicted by current-generation hadronic interaction models. This result can be interpreted as a deficit of muons in simulations as a lighter mass composition has been established from $X_{\mathrm{max}}$ measurements. This muon deficit was already observed in previous analyses of the Auger Collaboration and is confirmed using hybrid events that include radio measurements for the first time.

## I. INTRODUCTION

Ultra-high-energy cosmic rays can only be observed indirectly through extensive air showers initiated in the Earth's atmosphere. Their mass composition can be inferred from certain shower observables, such as the depth of the shower maximum, $X_{\mathrm{max}}$, and the "muon number," $N_\mu$, defined as the number of muons in the air shower detected at ground level. The muon number increases nearly linearly with the cosmic-ray energy, $E$, and with the mass number of the cosmic ray, $A$, i.e., $N_\mu = E^\beta A^{1-\beta}$ where $\beta \simeq 0.9$. The interpretation of the measured muon number in data relies on the comparison with predictions made by full Monte Carlo air-shower simulations based on hadronic interaction models. Previous studies conducted at the Pierre Auger Observatory have consistently observed more muons in data than predicted by current hadronic interaction models [1–3], while other experiments, such as Yakutsk, do not report a significant discrepancy [4]. A broader overview of results from nine air-shower experiments is given in Ref [5], highlighting the need for further investigation and novel approaches. Possible reasons for this "muon puzzle" and its connection to the Large Hadron Collider [6] remain an active area of research [7]. Recent work using the EPOS.LHC-R hadronic interaction model indicates an increase in the predicted muon number, which reduces the discrepancy with observations [8].

The Pierre Auger Observatory is the world's largest observatory for the detection of cosmic rays, covering an area of 3000 km$^2$ located in the province of Mendoza, Argentina. Its hybrid design enables the study of air showers over a wide energy range, from $10^{17}$ eV to beyond $10^{20}$ eV, using complementary detection techniques. The baseline detector systems (Auger Phase I) include the surface detector (SD) [9], an array of water-Cherenkov detectors arranged in triangular grids with three different spacings: 1500 m (SD-1500), 750 m (SD-750), and 433 m (SD-433), and the fluorescence detector (FD) [10] overlooking the SD array from four sites. These are complemented by dedicated enhancements targeting specific shower components such as the Underground Muon Detector (UMD) [11] and the Auger engineering radio array (AERA) [12]. As part of the AugerPrime upgrade [13], the SD has recently been enhanced with new detector components, including the surface scintillator detector [14] and the radio detector [15]. Furthermore, the SD station electronics have been upgraded.

The Pierre Auger Observatory has already measured the muon content in different energy ranges using data from the SD-1500 array combined with the FD [1,2], as well as from the SD-750 array and the UMD [3]. The potential of combining radio and muon measurements was demonstrated in simulations [16]. In this study, we present a novel measurement of the muon content for inclined air showers using hybrid events recorded by the SD-1500 array and AERA in coincidence.

AERA consists of 153 radio detector stations distributed over an area of 17 km$^2$ and is located in the northwestern part of the SD. It was deployed in three phases, gradually increasing its coverage and station spacing. The first 24 stations (AERA phase I) were installed in 2011 on a 144m triangular grid, covering 0.4 km$^2$. In 2013, an additional 100 stations (AERA phase II) were deployed with a larger spacing of 250m to 375m, expanding the array to 6 km$^2$. Finally, in 2015, the last 29 stations (AERA phase III) were added with a spacing of up to 750m, completing the current layout. A map of the individual deployment phases of AERA is presented in Fig. 1. Due to its small instrumented area, phase I is not suitable for reconstructing inclined air showers, so this analysis uses the data starting with phase

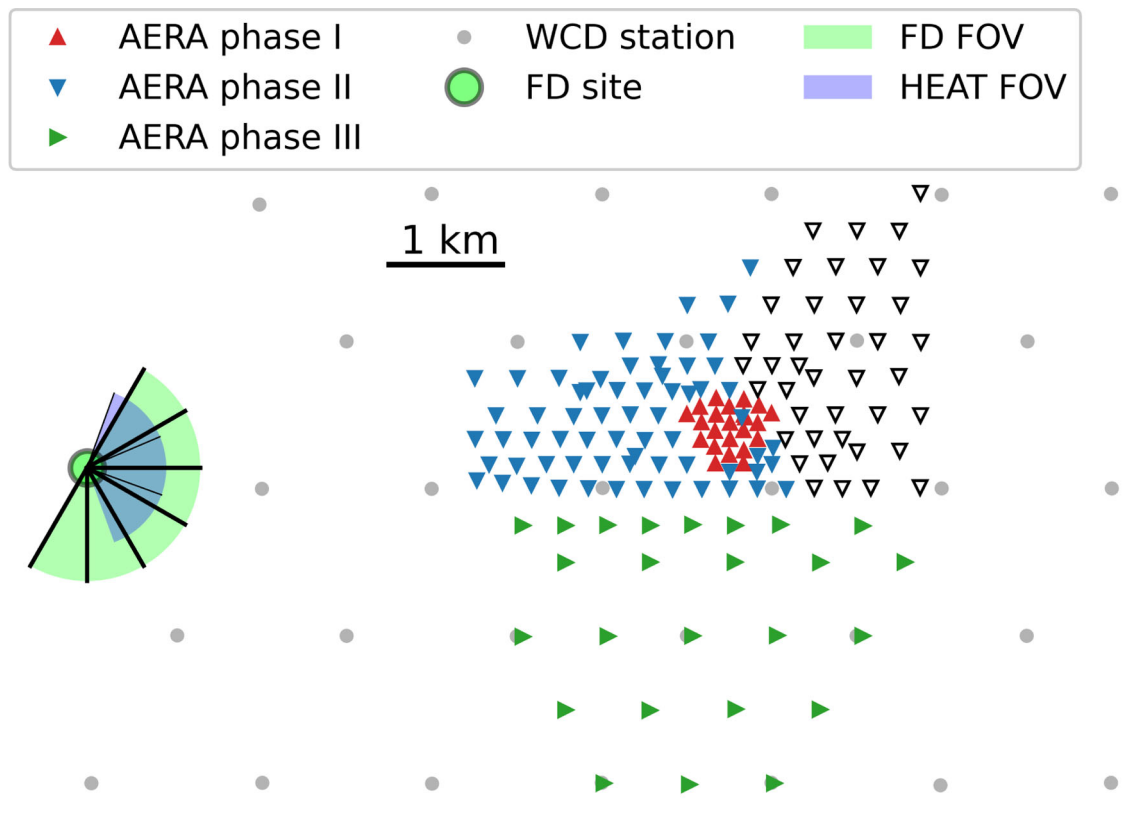

FIG. 1. Schematic map of AERA. The orientations of the triangles indicate the three deployment phases, empty triangles represent stations that cannot receive an external trigger and are not used in this analysis. Only the water-Cherenkov stations with a grid spacing of 1500 m are shown.

II. Only radio detector stations that can provide data upon an external trigger are used, amounting to 76 stations for phase II and 105 stations for phase III.

For inclined air showers with zenith angles greater than 60°, the electromagnetic component of the air shower is largely absorbed in the atmosphere and predominantly muons are detected by particle detectors on the ground. The radio emission, arising from the electromagnetic component of the air shower, is well understood and unaffected by atmospheric absorption or scattering, making it a robust tool for energy estimation [17,18]. This radio emission originates from two distinct mechanisms: the geomagnetic effect, which results from the deflection of electrons and positrons in the Earth's magnetic field, and the charge-excess effect, caused by a net negative charge buildup in the shower front [19]. Using a radio detector instead of the FD for the energy estimation has the benefit of a duty cycle of almost 100% whereas the FD has an uptime of only ∼15%. Furthermore, the geometric phase space for high-quality events reconstructed with the FD is small for inclined showers, as a large fraction of air showers have their $X_{\max}$ outside of the field of view of the telescopes. Such a selection is not needed with a radio detector, hence one can collect data more efficiently.

Currently, the analysis is constrained by low statistics due to the small area of AERA and the high energy threshold of 4 EeV required for the SD-1500 to operate at full efficiency. Therefore, this study serves as a proof of concept, demonstrating the feasibility of the proposed measurement technique. Consequently, we focus on the estimators for muon content and energy without converting them into high-level physical quantities of the air showers. This will be addressed in the long term by a higher-statistics analysis combining the AugerPrime Radio Detector and the SD as well as AERA with the SD stations deployed on the 750m grid at energies below 4 EeV.

## II. RECONSTRUCTION METHODS FOR INCLINED AIR SHOWERS

For the SD reconstruction, we use the well-established method for inclined showers described in Ref. [20], which is fully efficient for primary energies above 4 EeV. Inclined showers are characterized by elongated, asymmetrical footprints on the ground due to the long path the particles take through the atmosphere, causing significant geomagnetic deflections and extensive lateral spread. To good approximation, the shape of the muon distribution on the ground is found to be independent of the primary particle type, energy, and hadronic interaction model used in simulations. Differences manifest primarily in an overall normalization of the muon distribution. Therefore, two-dimensional reference maps of the lateral muon distribution on the ground, $\rho_{\mu,19}(\vec{r};\theta,\phi)$, are generated using 1800 proton showers simulated at an energy of $10^{19}$ eV and with QGSJet II-03 [21] as the hadronic interaction model. Here, $\vec{r}$ denotes the position on the ground relative to the shower core, while $\theta$ and $\phi$ are the zenith and azimuth angles of the primary particle, respectively.

The measured SD signals are corrected for the deposited energy by the residual electromagnetic component to obtain the muon density, $\rho_{\mu}(\vec{r};\theta,\phi,E)$, at each station. In the reconstruction, the reference maps are rescaled to match the measured signals of the SD stations, i.e.,

$$\rho_{\mu}(\vec{r};\theta,\phi,E) = N_{19}\rho_{\mu,19}(\vec{r};\theta,\phi). \tag{1}$$

The rescaling factor, $N_{19}$, serves as a relative measure of the muon content compared to the reference model. Using hybrid events reconstructed by both the FD and SD, $N_{19}$ correlates with the energy determined by the FD. Although it has some composition sensitivity, the data-driven calibration corrects the composition bias on average and thus provides an energy estimator for the cosmic ray [1].

For the radio signal, the signal distribution on the ground is described with a model specifically made for inclined air showers with zenith angles above 65° [22]. It was initially developed for the AugerPrime Radio Detector such that the application to AERA data requires validation. The model analytically computes the dominant geomagnetic radio emission based on the reconstructed shower geometry and the measured signal polarization. In this work, we use the shower direction reconstructed from the SD for this calculation. By effectively removing the asymmetry introduced by the charge-excess component, the remaining distribution of the geomagnetic radio emission can be described with a rotationally symmetric lateral distribution function (LDF). The LDF is modeled as the sum of a sigmoid function and a Gaussian-like function with a varying exponent outside the Cherenkov ring, capturing the transition between the dense radio signal near the Cherenkov radius and the more extended emission at larger distances. We will later use the position of the Gaussian-like function as an estimator for the Cherenkov radius. Integrating the LDF over the whole footprint yields the total radiation energy. After applying corrections for air density and geomagnetic angle, we obtain the "corrected radiation energy," $S_{\mathrm{rad}}$, which is directly related to the energy of the electromagnetic particle cascade, $E_{\mathrm{EM}} \propto \sqrt{S_{\mathrm{rad}}}$ [23].

An example event with a zenith angle of $\sim 78°$, arriving from $\sim 23°$ north of west, is shown in Fig. 2, illustrating both reconstruction methods. The SD event consists of 12 signal stations and is reconstructed with an energy of 5(1) EeV. In AERA, 59 stations recorded a signal well above the noise level, i.e., with a signal-to-noise ratio above 10, where the signal is defined as the square of the maximum of the Hilbert envelope of the electric field, and the noise as the square of the rms in a noise window. The upper panel of the figure displays the lateral distribution of signals in the SD, with the shaded band representing the predicted signal strength range for different positions in the shower plane, based on the corresponding reference map. The lower panel shows the distribution of the geomagnetic part of the radio signal, including the final fit of the lateral signal distribution model along with its individual component functions.

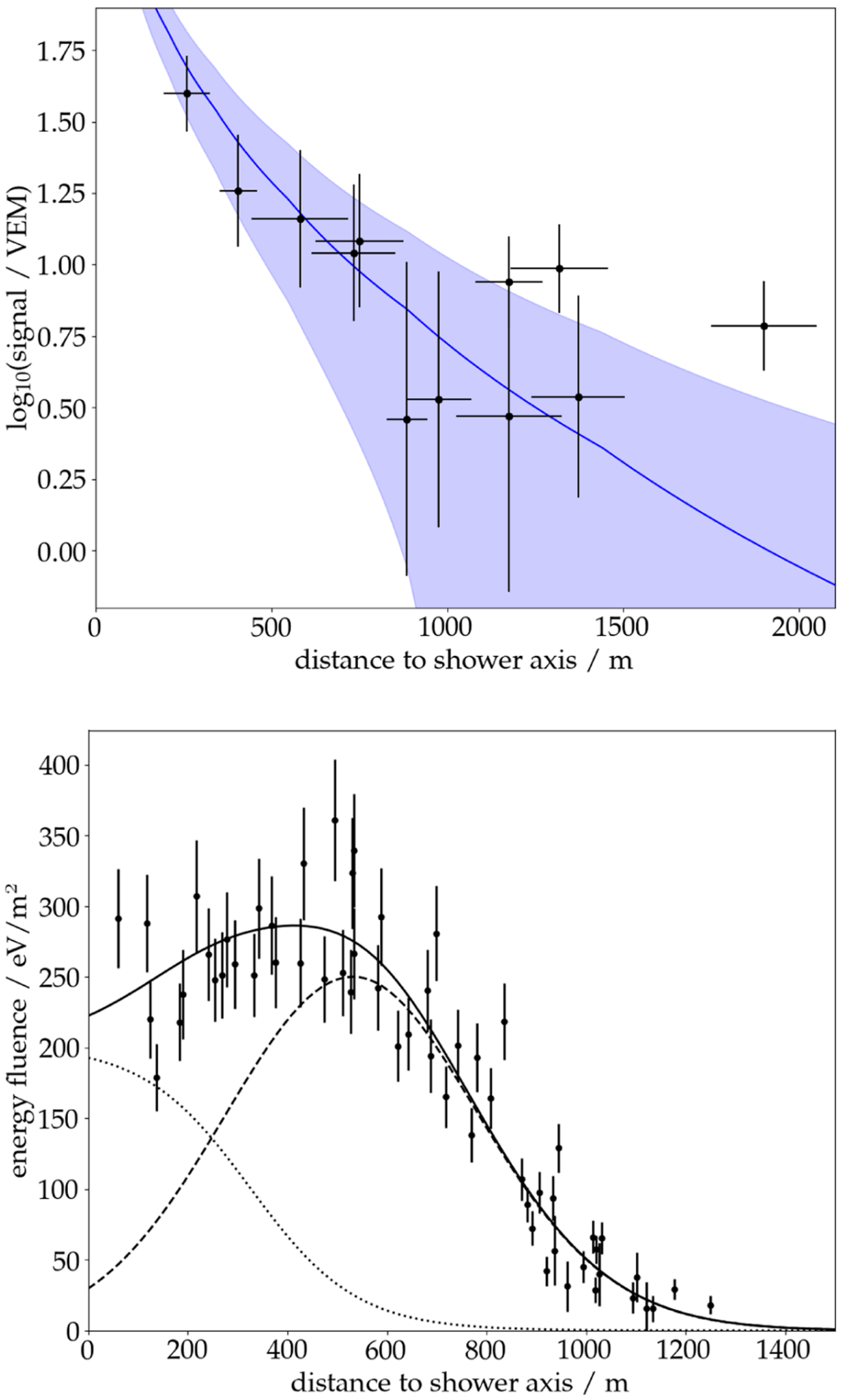

FIG. 2. Lateral signal distributions for an inclined air shower detected by the SD (top) and AERA (bottom). In the top panel, the solid line represents the expected signal strength predicted by the reference map, averaged over all polar angles in the shower plane. The blue band indicates the variation of the expected signal within the plane. The bottom panel shows the signal distribution of the geomagnetic radio emission together with the fitted LDF, composed of a Gaussian-like component (dashed line) and a sigmoid component (dotted line). The Cherenkov radius is determined as 531 m.

For the present work, data are presented as a function of $N_{19}$ and $S_{\mathrm{rad}}$, while events are selected in electromagnetic energy according to the conversion described in Ref [22].

## III. VALIDATION OF THE RADIO LDF MODEL

As the radio LDF model for inclined showers is applied to AERA data for the first time, we test its performance using a dedicated set of simulations covering the full phase space of possible event detections. The geometry and energy are sampled randomly following a uniform distribution in logarithmic energy and $\sin^2\theta$. Showers were generated with energies between 2 EeV and 40 EeV and zenith angles between 58° and 82°. The core positions were randomized such that a sufficient number of antennas is expected to be within a maximum of three Cherenkov radii (which range from approximately 200m in the shower plane at 60° to over 700 m at 80° zenith angle). This selection allows the shower core to be located outside the instrumented area of AERA. We simulate more than 1000 air showers with CoREAS [24] using QGSJet II-04 [25] as hadronic interaction model and proton and iron nuclei as primary particles.

The simulations are reconstructed including a realistic detector simulation and the addition of measured environmental noise from randomly selected timestamps. As the analysis is performed solely on estimator level we use the energy calibration of Ref. [22] to convert the Monte Carlo-true energy of the electromagnetic cascade to its corresponding corrected radiation energy for the validation of the LDF model.

We now apply a high-quality event selection. The SD reconstruction is used as an input for the radio one, therefore, we require at least five triggered stations in the SD array and we further require that all the six stations surrounding the station closest to the impact point of the shower on the ground are operational at the time of the event. This selection yields a bias-free energy reconstruction estimated from the number of muons for events with a primary energy above the full efficiency threshold of 4 EeV [20].

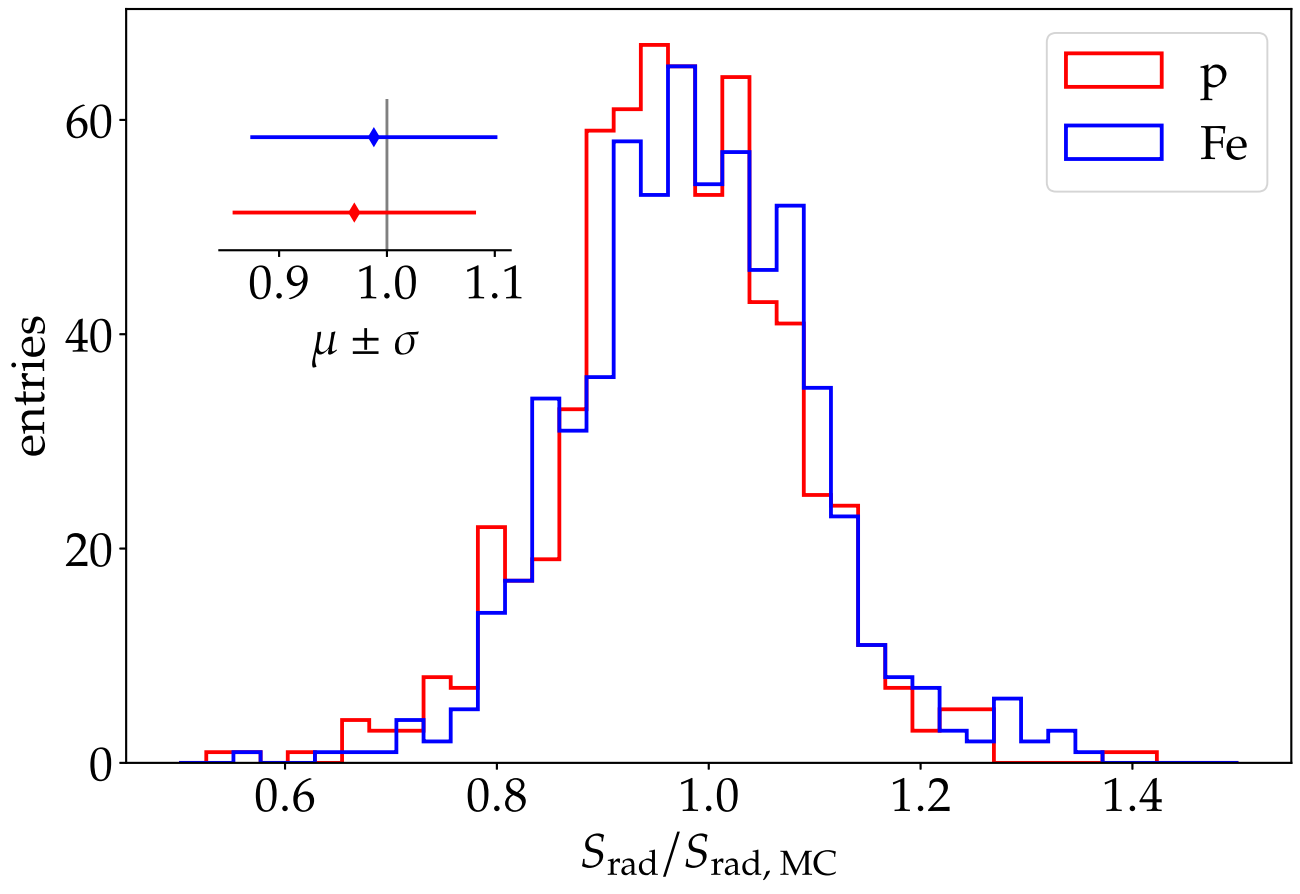

FIG. 3. Histogram showing the reconstruction accuracy of the corrected radiation energy for the subset of high-quality simulated events. The inset visualizes mean and standard deviation for different primaries.

To ensure high-quality fits of the radio LDF, several selection criteria are applied. Events must have signals in at least five stations, with at least one of them located within the Cherenkov ring (cf. Fig. 2, bottom). Additionally, we require the LDF fit to yield a reduced $\chi^2$ below 5. As radio detection provides direct access to the electromagnetic energy, not the primary energy, we require that $E_{\mathrm{EM}}$ is above 4 EeV. This threshold guarantees that the primary energy exceeds the full efficiency threshold for the SD reconstruction.

It is known that the fitting procedure tends to underestimate the uncertainty of the reconstructed energy [26]. To account for this, we increase the uncertainty of the reconstructed $S_{\mathrm{rad}}$ by 10% in quadrature, ensuring that the pull distribution, i.e., the number of standard deviations by which the observed values deviate from the expected values, more closely resembles a normal distribution. Occasionally, the reconstruction also predicts very large uncertainties. Thus, we only select events with a relative uncertainty on the reconstructed $\sqrt{S_{\mathrm{rad}}}$ below 20%.

The resulting reconstruction performance for the energy estimator is shown in Fig. 3. We observe a mean underestimation of 3% for protons and 1% for iron primaries. Thus, the primary-dependent bias is considered negligible. The remaining bias is likely attributed to signal processing, such as the removal of radio frequency interference. Since this bias is small, it will not be further investigated here. For simplicity, we take this residual bias as 2%, corresponding to the average of the observed biases for protons and iron nuclei, and account for it as a systematic uncertainty. The spread of 11% in $S_{\mathrm{rad}}$ observed for both primaries translates into an energy resolution of 5.5%. No significant dependence on the zenith angle is observed.

## IV. MUON CONTENT IN INCLINED AIR SHOWERS

In this analysis, we examine the AERA data recorded between June 26, 2013 (start of AERA phase II) until August 10, 2023, the last event reconstructed with the Auger Phase I SD. We first characterize the phase space for hybrid SD-AERA event detection by requiring only a minimal selection for the reconstructed events. A successful geometry reconstruction by both the SD and AERA is required, and the zenith angle reconstructed by the SD must be between 65° and 80°. This zenith range is chosen because the LDF model of the radio emission is only applicable for zenith angles above 65°; it also ensures that the SD signal is predominantly due to muons. For very inclined showers above 80°, the performance of the SD reconstruction deteriorates. Additionally, events must not be recorded during thunderstorm conditions as identified by the electric field mills or periods where no reliable electric-field-mill data are available [27]. This selection results in 2103 candidate events.

The radio directional reconstruction is performed using a spherical wavefront fit to the arrival time of the radio pulse at the signal stations [27] since the shower develops far away from the observer positions on the ground. The fit reconstructs a “hypothetical” point source from which the radio signal appears to originate, which is then combined with the shower core as determined by the SD to obtain the shower axis. In general, the reconstructed directions from AERA and the SD agree well. For 89.5% of events, the opening angle, defined as the angle between the reconstructed shower directions from the SD and AERA is less than 2°, only 6.1% of events have an opening angle larger than 5°. These larger discrepancies mainly occur in events with only a few radio signal stations. Upon closer examination, these outliers often arise from radio events where strong noise pulses are mistakenly identified as air-shower signals, or where the radio footprint lies mostly outside the array, resulting in only weak signals at stations near its edge. In the following steps of the analysis, the shower direction reconstructed from the SD is used.

As expected, more events are observed coming from the south than from the north, where the larger angle to Earth’s magnetic field results in stronger geomagnetic radio emission. More events are detected at larger inclinations as the size of the radio footprint increases with zenith angle [28]. This allows the reconstruction of events whose shower cores fall outside the instrumented area of AERA. The spatial distribution of shower core positions, as reconstructed by the SD, is shown in Fig. 4. To determine the density of these reconstructed positions, a kernel density estimation is applied using a normal distribution as the kernel and a spread derived from Scott’s rule [29]. While most events have their shower core within or close to AERA, very inclined showers can still be reconstructed when their impact points fall well outside.

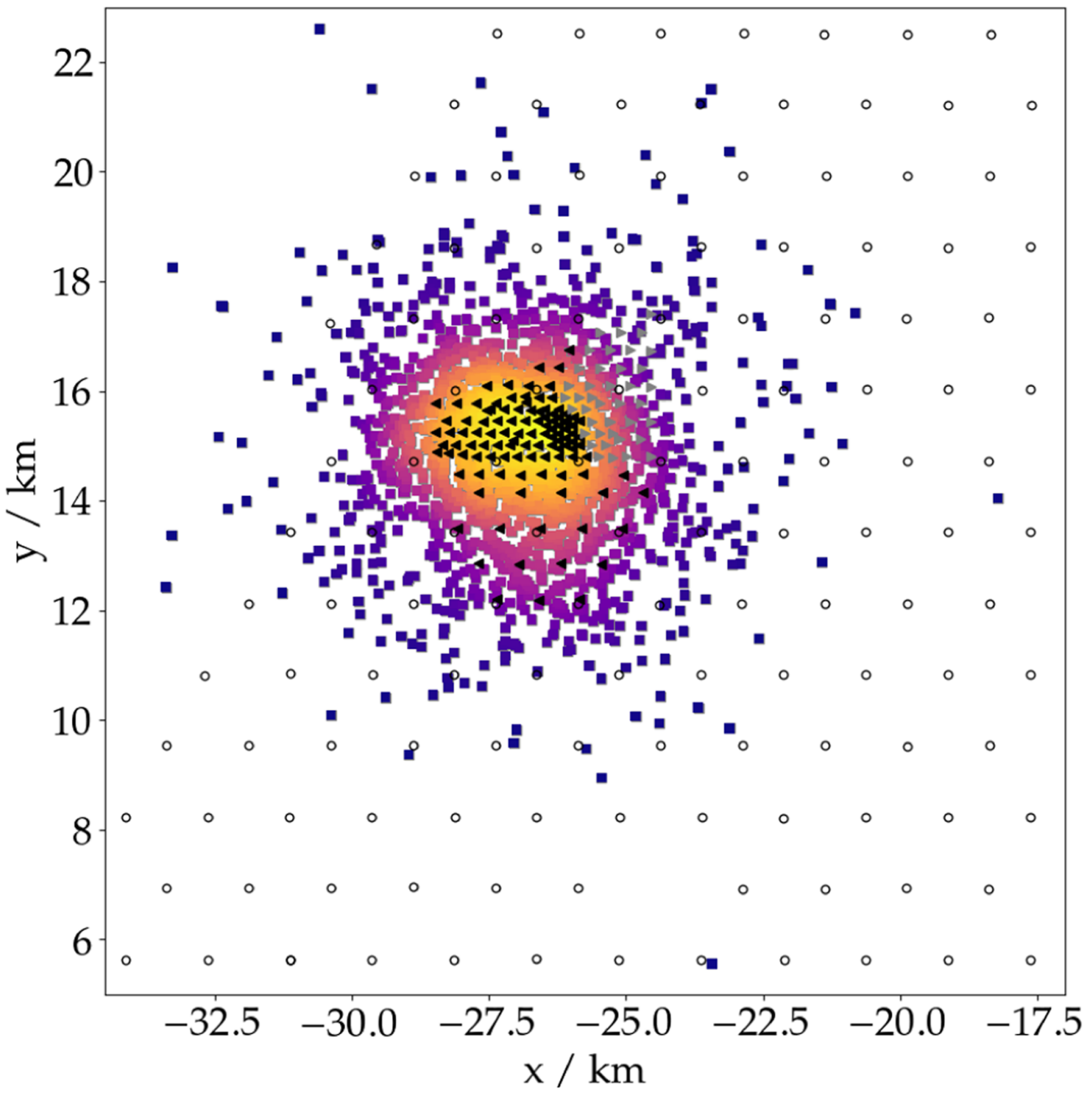

FIG. 4. Distributions of the impact points of the 2103 candidate air showers determined with the SD, shown as square markers. The color indicates the density as determined by a KDE, with low densities shown in dark purple and high densities in bright yellow. Circles mark the positions of the SD stations, and triangles indicate the locations of the AERA stations.

We apply the high-quality event selection from the previous section to the measured events. This selection yields 40 high-quality hybrid events with energies reconstructed by the SD between 3.4(7) EeV and 12.6(12) EeV. The number of events remaining after each cut is detailed in Table I. The most restrictive cut is the minimum $E_{\mathrm{EM}}$ threshold of 4 EeV as reconstructed by AERA, which corresponds to the threshold for full efficiency of the SD for inclined events. Given the high energy of the selected events, the large footprint of the radio emission on ground for inclined showers, and the dense layout of AERA, the hybrid detection with SD and AERA is also expected to be fully efficient.

An intensive study of systematic and event-dependent uncertainties for AERA has already been performed in Ref. [18]. The event-by-event uncertainties are dominated by the temperature dependence of the electronic signal chain and the uncertainty of the directional-dependent antenna response pattern. They manifest as an additional scatter in the measured data, hence, the uncertainty of the $S_{\mathrm{rad}}$ reconstruction is increased by 12.8%. Techniques to mitigate these effects, e.g., a relative antenna calibration using a drone-mounted reference antenna [30], are currently under development. The dominating uncertainties on the absolute scale are given by those on the absolute scale of the antenna response pattern (25% for $S_{\mathrm{rad}}$) and the electronic signal chain (12% for $S_{\mathrm{rad}}$). In addition, we take the remaining bias of the reconstruction, cf. Sec. III, into account as a systematic uncertainty of 2%. In total, this results in a systematic uncertainty of 27.8% on the absolute scale of the $S_{\mathrm{rad}}$ reconstruction. Note that this translates to half of that value on $E_{\mathrm{EM}}$. Additionally, this uncertainty can be significantly reduced in future analyses utilizing the continuously monitored sidereal modulation of the diffuse Galactic radio emission [31]. The systematic uncertainty on the SD reconstruction is quantified as 10%, obtained from the uncertainty of the detector response [32]. An overview of all event-by-event and absolute scale uncertainties is given in Table II.

The measured muon content in data is presented in Fig. 5 as a function of $\sqrt{S_{\mathrm{rad}}}$. The profile shows an expected increase of the number of muons for increasing values of $\sqrt{S_{\mathrm{rad}}}$, i.e., with increasing energy. The profile using equal logarithmic bins appears to flatten for events with $\sqrt{S_{\mathrm{rad}}/\mathrm{GeV}} \lesssim 1$. However, we have checked that this is not a systematic turn-off but rather a statistical fluctuation due to the low number of events in this range.

TABLE I. Number of events after each cut starting with 4067 reconstructed events. The first group of cuts corresponds to the basic selection step to describe the overall dataset, followed by the high-quality selection criteria for the SD and AERA reconstruction, respectively.

| Selection criterion | Events remaining |
|---|---|
| $65° \leq \theta_{\mathrm{SD}} \leq 80°$ | 2360 |
| No thunderstorm conditions | 2103 |
| Number of triggered SD stations $\geq 5$ | 1205 |
| Full hexagon of active stations | 974 |
| $E_{\mathrm{EM}} > 4$ EeV | 133 |
| Station within Cherenkov radius | 61 |
| Number of AERA signal stations $\geq 5$ | 51 |
| Reduced $\chi^2$ of LDF fit $< 5$ | 42 |
| Relative $\sqrt{S_{\mathrm{rad}}}$ uncertainty $< 0.2$ | 40 |

TABLE II. Overview of systematic uncertainties categorized into event-by-event and absolute scale uncertainties for $N_{19}$ and $S_{\mathrm{rad}}$. Event-by-event uncertainties are only reported for $S_{\mathrm{rad}}$ and added in quadrature to the fit uncertainty.

| Source of uncertainty | Value % |
|---|---|
| *Event-by-event uncertainties* | 12.8 |
| Temperature dependence | 8 |
| Angular dependence of antenna response | 10 |
| *Absolute scale uncertainties* | |
| $S_{\mathrm{rad}}$ *total* | 27.8 |
| Antenna response pattern | 25 |
| Analog signal chain | 12 |
| LDF model | 2 |
| $N_{19}$ *total* | 10 |
| Detector response | 10 |

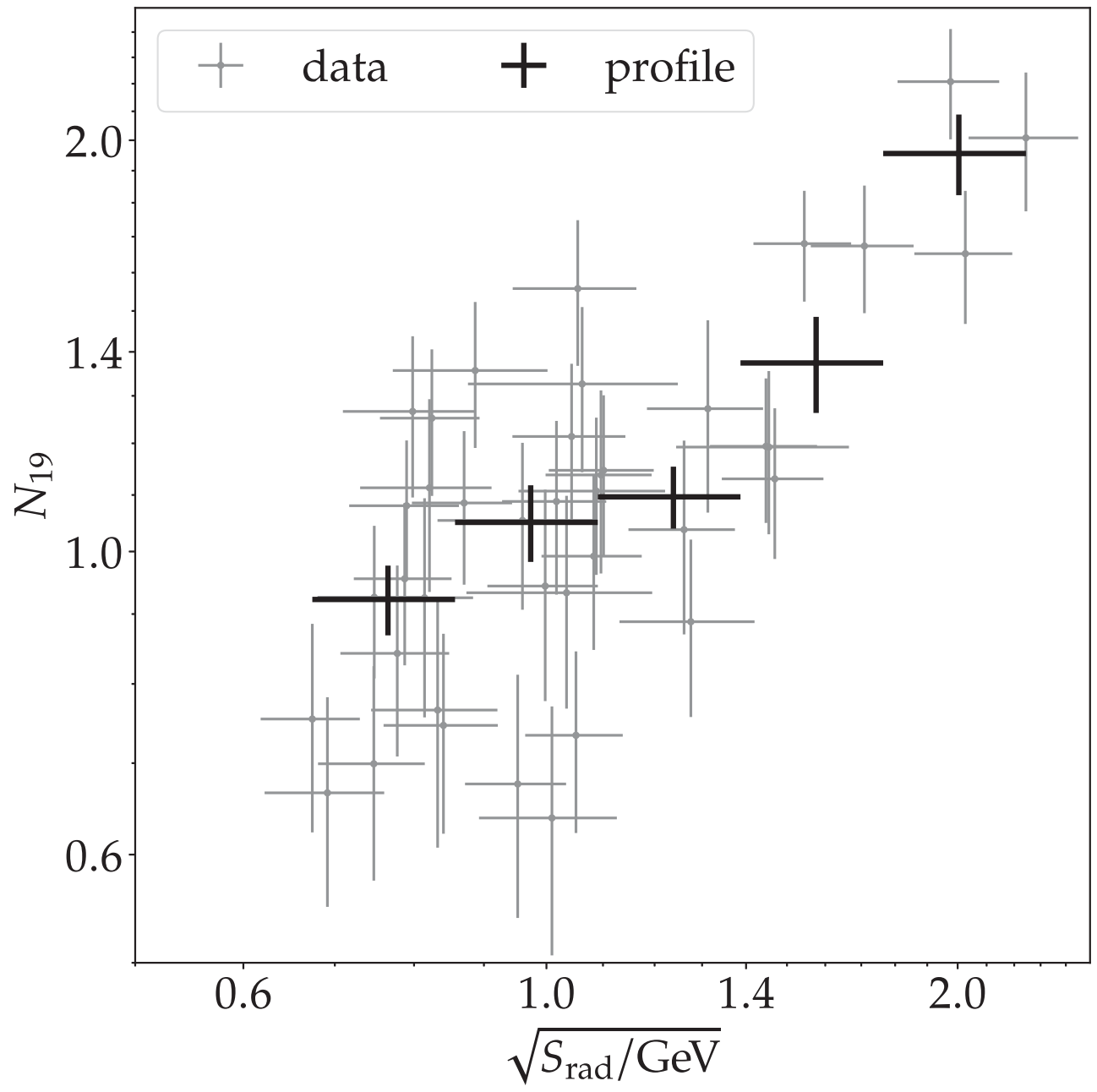

FIG. 5. Measured muon content estimator, $N_{19}$, as a function of the energy estimator, $\sqrt{S_{\mathrm{rad}}}$. For each measured event, the reconstructed estimators and their uncertainties are shown by the gray data points. The black profile denotes the average for each energy bin, the $y$ uncertainty is given by the uncertainty of the mean.

For an interpretation of the measured muon content, we compute predictions of the muon content from simulations. We utilize over 100 000 inclined air showers simulated with CORSIKA [33] using QGSJet II-04, EPOS-LHC [34], and Sibyll 2.3d [35] as high-energy hadronic interaction models. The simulations use protons and iron nuclei as primaries with energies between $10^{18.4}$ eV and $10^{19.6}$ eV. The electromagnetic energy of each air shower is calculated as the sum of the energy deposited by all electromagnetic particles, which is then converted to the corresponding $S_{\mathrm{rad}}$ based on Ref. [22]. Each simulated air shower is reconstructed using the standard Auger analysis framework [36,37] to obtain $N_{19}$. Finally, we fit $N_{19}$ as a function of $\sqrt{S_{\mathrm{rad}}}$ using a power-law function, i.e., $N_{19} = a\sqrt{S_{\mathrm{rad}}}^{b}$. The resulting fit parameters are listed in Table III.

TABLE III. Fit parameters of the relation $N_{19} = a\sqrt{S_{\mathrm{rad}}}^{b}$ for different hadronic interaction models and primary particles.

| Model | Primary | $a$ | $b$ |
|---|---|---|---|
| SIBYLL 2.3d | Proton | 0.7108 | 0.9169 |
| | Iron | 0.9987 | 0.9225 |
| EPOS-LHC | Proton | 0.6926 | 0.9293 |
| | Iron | 0.9722 | 0.9266 |
| QGSJetII-04 | Proton | 0.6605 | 0.9435 |
| | Iron | 0.9587 | 0.9203 |

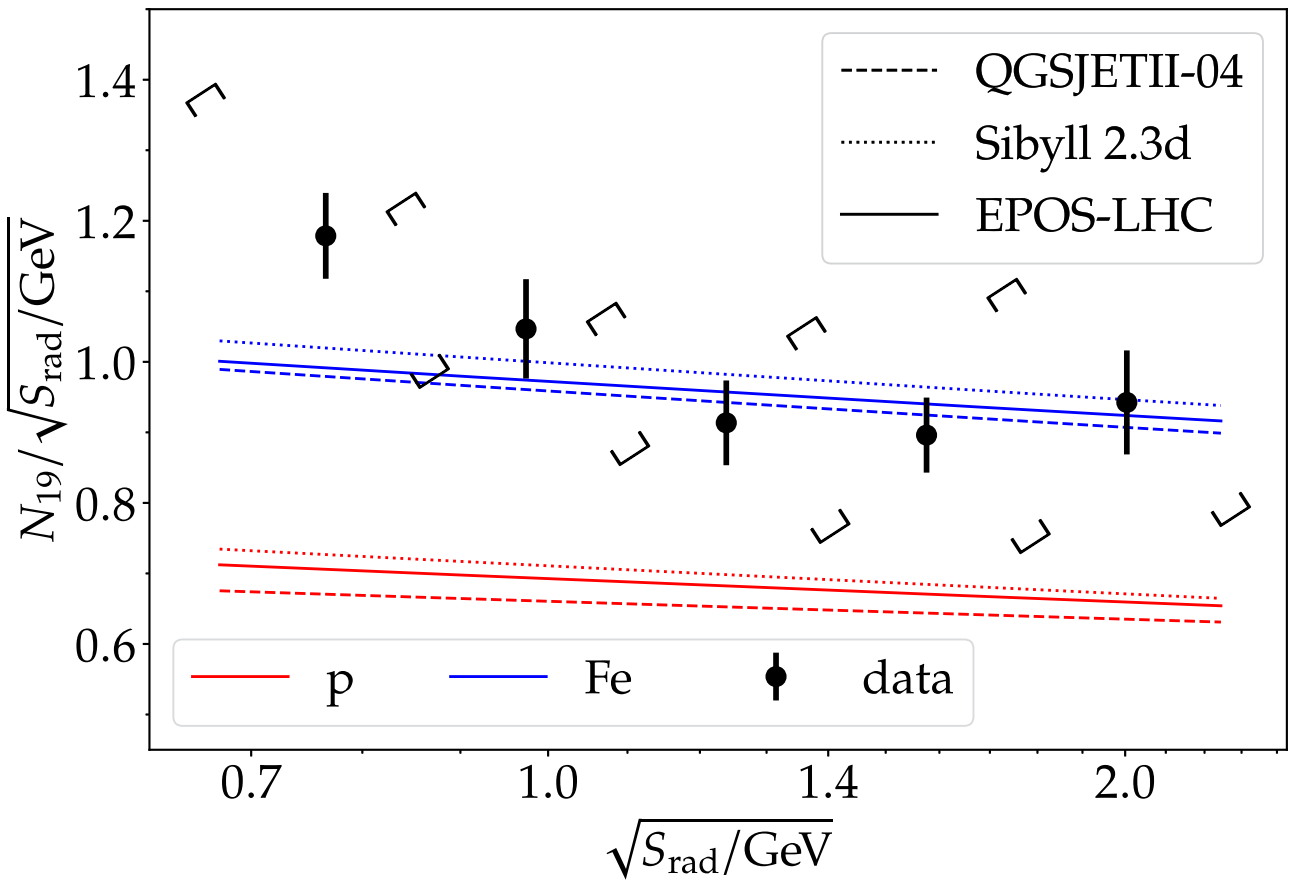

FIG. 6. Normalized muon content as a function of energy estimator. The predictions for different hadronic interaction models are denoted by the colored lines for protons and iron primaries. Vertical error bars indicate the statistical uncertainty of each data point, while square brackets indicate the systematic uncertainty of the measurement as listed in Table II. The diagonal offsets represent the correlated effect of systematic shifts in the energy estimator.

The average muon content normalized by the energy estimator is shown in Fig. 6. Dividing $N_{19}$ by $\sqrt{S_{\mathrm{rad}}}$ compensates for the expected energy dependence of the muon number (cf. Sec. I) and enables a better comparison with model predictions. Note that $N_{19}/\sqrt{S_{\mathrm{rad}}} \approx 1$ for iron primaries is a coincidence, as $S_{\mathrm{rad}} = 1$ GeV corresponds to an energy below $10^{19}$ eV. It is compatible with the prediction of hadronic interaction models for iron nuclei. Without knowledge of the radio energy scale, i.e., how $\sqrt{S_{\mathrm{rad}}}$ relates to the primary energy as reconstructed by the FD, a precise prediction of the expected muon content as a function of $\sqrt{S_{\mathrm{rad}}}$ cannot be given. Although in principle the mass composition can be derived from $X_{\mathrm{max}}$ measurements of the Auger FD, the events we analyze here do not have a FD counterpart. From other analyses in the energy range of this work, the mean atomic mass number is found to be between proton and nitrogen [38]. While the current data match the prediction for an iron primary, a lighter composition would also be compatible given the current systematic uncertainties. The presented result, based on the novel radio detection technique, is in broad agreement with previous Auger analyses using different detector combinations [1–3], in which a deficit of muons in simulations was reported.

## V. SUMMARY AND CONCLUSIONS

We presented a first estimate of the muon content of inclined air showers using hybrid measurements combining radio and particle detection. This serves as a proof of concept for future analyses with hybrid radio and particle events. We find a muon content in data that is compatible with the prediction of hadronic interaction models for iron-induced air showers even though the composition is expected to be between proton and nitrogen. Overall, this result is in broad agreement with previous Auger analyses using different detector combinations [1–3], which reported a deficit of muons in simulations, and demonstrates the potential of hybrid radio-particle measurements for studying the muon content of air showers.

Currently, the analysis is limited by the low statistics of 40 high-quality events originating from the small area of AERA and the high energy threshold of 4 EeV required for the reconstruction with the 1500 m SD array. Even without the high-quality selection for the radio events, this detector combination would yield at most 133 events above the full-efficiency threshold, cf. Table I, insufficient for high-impact results. Instead, we plan to apply the same method to different detector combinations in future publications. An adaption of the inclined reconstruction technique used for the 1500 m SD array is currently being developed for the 750 m array, which will considerably reduce the energy threshold and, therefore, the collection of higher statistics at energies below 4 EeV. This larger dataset will also enable tests of potential systematic effects related to the orientation of the shower axis with respect to the geomagnetic field, which influences the shape of the lateral distribution function and the mean muon energy. With the AugerPrime Radio Detector recently completed, this analysis can also be extended to the highest energies to allow for in-depth tests of hadronic interaction models with large statistics [15,39].

## ACKNOWLEDGMENTS

The computations were partially carried out on the PLEIADES cluster at the University of Wuppertal, which was supported by the Deutsche Forschungsgemeinschaft (DFG, Grant No. INST 218/78-1 FUGG) and the Bundesministerium für Bildung und Forschung (BMBF). The successful installation, commissioning, and operation of the Pierre Auger Observatory would not have been possible without the strong commitment and effort from the technical and administrative staff in Malargüe. We are very grateful to the following agencies and organizations for financial support: Argentina—Comisión Nacional de Energía Atómica; Agencia Nacional de Promoción Científica y Tecnológica (ANPCyT); Consejo Nacional de Investigaciones Científicas y Técnicas (CONICET); Gobierno de la Provincia de Mendoza; Municipalidad de Malargüe; NDM Holdings and Valle Las Leñas; in gratitude for their continuing cooperation over land access; Australia—the Australian Research Council; Belgium—Fonds de la Recherche Scientifique (FNRS); Research Foundation Flanders (FWO), Marie Curie Action of the European Union Grant No. 101107047; Brazil—Conselho Nacional de Desenvolvimento Científico e Tecnológico (CNPq); Financiadora de Estudos e Projetos (FINEP); Fundação de Amparo à Pesquisa do Estado de Rio de Janeiro (FAPERJ); São Paulo Research Foundation (FAPESP) Grants No. 2019/10151-2, No. 2010/07359-6, and No. 1999/05404-3; Ministério da Ciência, Tecnologia, Inovações e Comunicações (MCTIC); Czech Republic—GACR 24-13049S, CAS LQ100102401, MEYS LM2023032, CZ.02.1.01/0.0/0.0/16_013/0001402, CZ.02.1.01/0.0/0.0/18_046/0016010, CZ.02.1.01/0.0/0.0/17_049/0008422, and CZ.02.01.01/00/22_008/0004632; France—Centre de Calcul IN2P3/CNRS; Centre National de la Recherche Scientifique (CNRS); Conseil Régional Ile-de-France; Département Physique Nucléaire et Corpusculaire (PNC-IN2P3/CNRS); Département Sciences de l'Univers (SDU-INSU/CNRS); Institut Lagrange de Paris (ILP) Grant No. LABEX ANR-10-LABX-63 within the Investissements d'Avenir Programme Grant No. ANR-11-IDEX-0004-02; Germany—Bundesministerium für Bildung und Forschung (BMBF); Deutsche Forschungsgemeinschaft (DFG); Finanzministerium Baden-Württemberg; Helmholtz Alliance for Astroparticle Physics (HAP); Helmholtz-Gemeinschaft Deutscher Forschungszentren (HGF); Ministerium für Kultur und Wissenschaft des Landes Nordrhein-Westfalen; Ministerium für Wissenschaft, Forschung und Kunst des Landes Baden-Württemberg; Italy—Istituto Nazionale di Fisica Nucleare (INFN); Istituto Nazionale di Astrofisica (INAF); Ministero dell'Università e della Ricerca (MUR); CETEMPS Center of Excellence; Ministero degli Affari Esteri (MAE), ICSC Centro Nazionale di Ricerca in High Performance Computing, Big Data and Quantum Computing, funded by European Union NextGenerationEU, reference code CN_00000013; México—Consejo Nacional de Ciencia y Tecnología (CONACYT) No. 167733; Universidad Nacional Autónoma de México (UNAM); PAPIIT DGAPA-UNAM; The Netherlands—Ministry of Education, Culture and Science; Netherlands Organisation for Scientific Research (NWO); Dutch national e-infrastructure with the support of SURF Cooperative; Poland—Ministry of Education and Science, Grants No. DIR/WK/2018/11 and No. 2022/WK/12; National Science Centre, Grants No. 2016/22/M/ST9/00198, No. 2016/23/B/ST9/01635, No. 2020/39/B/ST9/01398, and No. 2022/45/B/ST9/02163; Portugal—Portuguese national funds and FEDER funds within Programa Operacional Factores de Competitividade through Fundação para a Ciência e a Tecnologia (COMPETE); Romania—Ministry of Research, Innovation and Digitization, CNCS-UEFISCDI, Contract No. 30N/2023 under Romanian National Core Program LAPLAS VII, Grant No. PN 23 21 01 02 and Project No. PN-III-P1-1.1-TE-2021-0924/TE57/2022, within PNCDI III; Slovenia—Slovenian Research Agency, Grants No. P1-0031, No. P1-0385, No. I0-0033, and No. N1-0111; Spain—Ministerio de Ciencia e Innovación/Agencia Estatal de Investigación (PID2019–105544 GB-I00, PID2022-140510NB-I00, and RYC2019-027017-I), Xunta de Galicia (CIGUS Network of Research Centers, Consolidación 2021 GRC GI-2033, ED431C-2021/22 and ED431F-2022/15), Junta de Andalucía (SOMM17/6104/UGR and P18-FR-4314), and the European Union (Marie Sklodowska-Curie 101065027 and ERDF); USA—Department of Energy, Contracts No. DE-AC02-07CH11359, No. DE-FR02-04ER41300, No. DE-FG02-99ER41107, and No. DE-SC0011689; National Science Foundation, Grant No. 0450696, and NSF-2013199; The Grainger Foundation; Marie Curie-IRSES/EPLANET; European Particle Physics Latin American Network; and UNESCO.

## DATA AVAILABILITY

The data that support the findings of this article are openly available [40].